\documentclass[aps,prl,twocolumn]{revtex4}
\usepackage{bm}
\usepackage{tabularx}
\usepackage{ascmac}
\usepackage{amsmath}
\usepackage{wrapfig}
\usepackage{graphicx}
\usepackage{float}
\usepackage[FIGBOTCAP]{subfigure}
\usepackage{color}

\bmdefine{\boldb}{b}
\bmdefine{\bolds}{s}
\bmdefine{\boldS}{S}
\bmdefine{\boldso}{so}
\bmdefine{\boldSO}{SO}
\bmdefine{\boldv}{v}
\bmdefine{\boldi}{i}
\bmdefine{\boldj}{j}
\bmdefine{\boldtau}{\tau}
\bmdefine{\boldsigma}{\sigma}
\bmdefine{\boldl}{l}
\bmdefine{\boldH}{H}
\bmdefine{\boldL}{L}
\bmdefine{\boldnabla}{\nabla}
\bmdefine{\boldlambda}{\lambda}
\bmdefine{\boldx}{x}
\bmdefine{\boldX}{X}
\bmdefine{\boldk}{k}
\bmdefine{\boldK}{K}
\bmdefine{\boldp}{p}
\bmdefine{\boldq}{q}
\bmdefine{\boldQ}{Q}
\bmdefine{\boldD}{D}
\bmdefine{\boldr}{r}
\bmdefine{\boldR}{R}
\bmdefine{\boldn}{n}
\bmdefine{\boldj}{j}
\bmdefine{\boldA}{A}
\bmdefine{\boldzero}{0}
\bmdefine{\boldone}{1}
\bmdefine{\boldtwo}{2}
\bmdefine{\boldthree}{3}
\bmdefine{\boldfour}{4}

\begin{document}


\title{
Weak localization of magnons 
in a disordered two-dimensional antiferromagnet
}


\author{Naoya Arakawa}
\email{naoya.arakawa@sci.toho-u.ac.jp} 
\author{Jun-ichiro Ohe}
\affiliation{
Department of Physics, Toho University, 
Funabashi, Chiba, 274-8510, Japan}


\begin{abstract}
We propose the weak localization of magnons in a disordered two-dimensional antiferromagnet. 
We derive the longitudinal thermal conductivity $\kappa_{xx}$ for magnons 
of a disordered Heisenberg antiferromagnet 
in the linear-response theory with the linear-spin-wave approximation. 
We show that 
the back scattering of magnons is enhanced critically by 
the particle-particle-type multiple impurity scattering. 
This back scattering causes 
a logarithmic suppression of $\kappa_{xx}$ with the length scale in two dimensions. 
We also argue a possible effect of inelastic scattering on 
the temperature dependence of $\kappa_{xx}$. 
This weak localization is useful to 
control turning the magnon thermal current on and off.

\end{abstract}

\date{\today}
\maketitle



\textit{Introduction.}
The Anderson localization 
is an impurity-induced localization of electrons~\cite{Anderson}. 
Its effects depend on the dimension of the system and 
the symmetry of the Hamiltonians~\cite{RG-4persons,Gorkov,Lee,Hikami}. 
The understanding has been advanced substantially by the theory in the weak-localization regime, 
where the effects of impurities can be treated 
as perturbation~\cite{Gorkov,Lee,Hikami,WeakLoc-review,Nagaoka}. 
For example, 
the weak-localization theory of a disordered two-dimensional electron system 
demonstrates the logarithmic temperature dependence of the resistivity, 
the negative magnetoresistance, 
and the antilocalization due to the spin-orbit coupling; 
those are experimentally confirmed~\cite{exp-logT,exp-NegativeMR,exp-SOC}. 
That theory also reveals 
the Anderson localization originates from the critical back scattering 
due to the multiple electron-electron scattering 
under time-reversal symmetry~\cite{WeakLoc-review}. 

Since the similar argument may be applicable to magnons, 
quasiparticles in a magnet, 
the weak localization of magnons 
has the potential for a new avenue in spintronics. 
Among several possibilities, 
antiferromagnets are suitable 
because global time-reversal symmetry holds 
and because 
even nondisordered antiferromagnets have several applications~\cite{Jungwirth}. 
(In contrast to electron systems, 
local time-reversal symmetry is broken in any magnets due to the magnetic ordering.) 
Then 
the knowledge for disordered antiferromagnets 
will be useful for others, such as disordered ferromagnets, 
which break global time-reversal symmetry. 
As well as antiferromagnets, 
ferromagnets are useful 
for carrying information and energy~\cite{Kajiwara,Uchida,Bauer}.

Despite the above potential, 
it is unclear how impurities affect magnon transport 
even in the weak-localization regime. 
In particular, 
the weak-localization theory of magnons under global time-reversal symmetry 
will be highly desirable 
because 
the previous theories~\cite{MagLoc1,MagLoc2,MagLoc3,MagLoc4} 
about the magnon localization 
analyze ferromagnetic cases, 
in which global time-reversal symmetry is broken. 
Although there is a previous theory~\cite{MagLoc-exception} about the magnon localization 
in an antiferromagnetic case, 
that does not study magnon transport.  
Since the existence of the back scattering is not sufficient to justify 
the localization, 
it is necessary to study magnon transport 
in disordered antiferromagnets. 
In particular, it is essential to clarify whether the weak localization 
occurs or not in the presence of global time-reversal symmetry 
without local time-reversal symmetry 
and how the weak localization of magnons is characterized by an observable quantity.

\begin{figure}[tb]
\includegraphics[width=68mm]{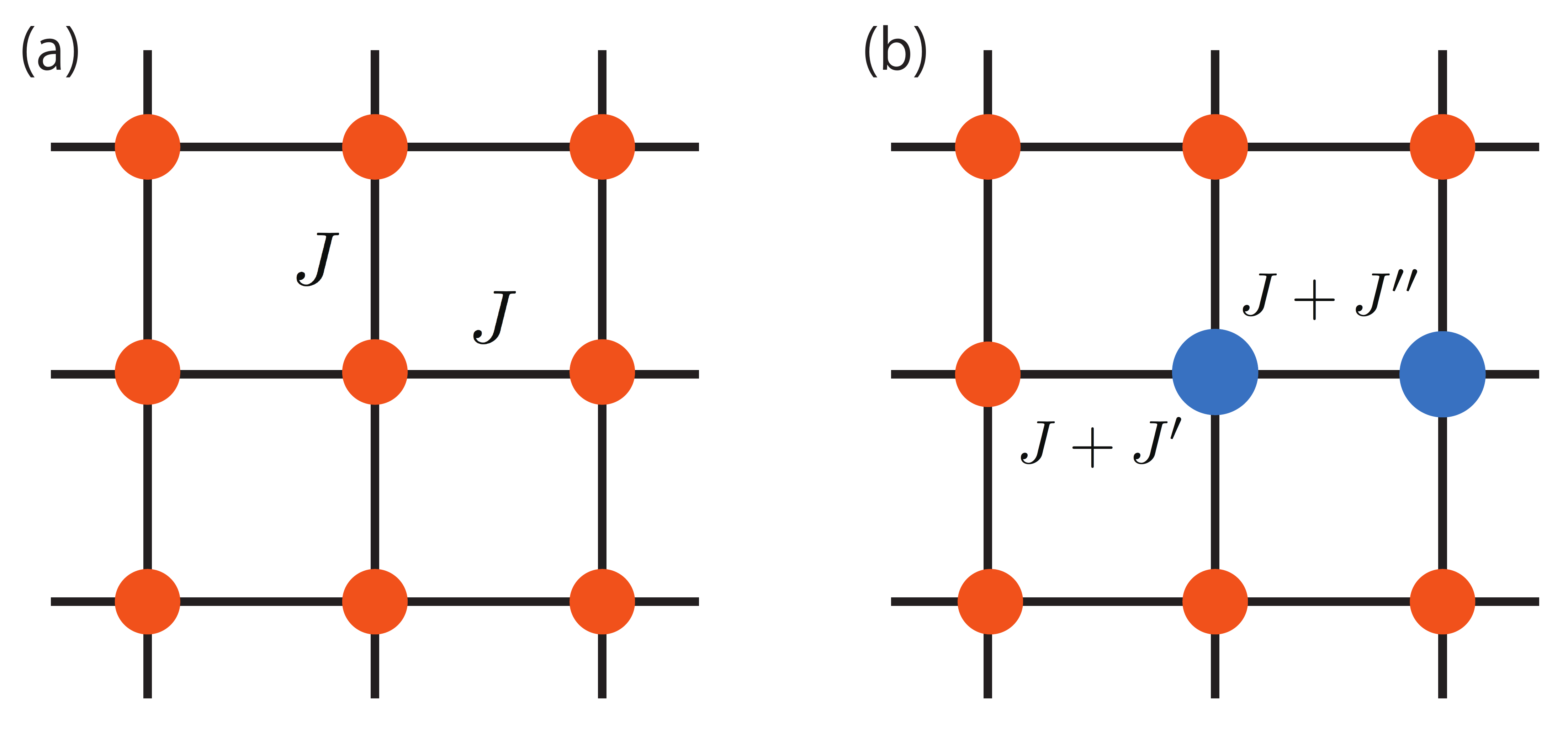}
\caption{
Schematic pictures of a lattice (a) without and (b) with disorder. 
An orange circle represents a magnetic ion, 
and a blue circle represents a different one. 
$J$, $J+J^{\prime}$, and $J+J^{\prime\prime}$ 
are the Heisenberg interactions between orange circles, 
between orange and blue circles, 
and between blue circles. 
}
\label{fig1}
\end{figure}

In this Rapid Communication 
we formulate the longitudinal thermal conductivity $\kappa_{xx}$ 
of magnons in a disordered Heisenberg antiferromagnet, 
and show disorder effects in the weak-localization regime. 
Our formulation is based on the linear-response theory~\cite{Luttinger,Oji-Streda,Tatara} 
with the linear-spin-wave approximation~\cite{Anderson-SW}. 
In our model, 
disorder is induced by 
partial substitution for magnetic ions [Fig. \ref{fig1}(b)], 
and its main effect is considered as changing the value of the Heisenberg interaction. 
We show that 
the particle-particle-type multiple impurity scattering of magnons 
causes the critical back scattering 
for any dimension and any spin quantum number $S$. 
Most importantly, 
this critical back scattering 
drastically suppresses the magnon thermal flow in two dimensions. 
We also argue a possible temperature dependence of $\kappa_{xx}$ 
in the presence of inelastic scattering. 
We finally discuss the validity of our theory 
and implications of experiments and theories. 
Throughout this paper we set $k_{\textrm{B}}=1$ and $\hbar=1$. 

\textit{Model.}
We begin to construct a model for a disordered antiferromagnet. 
Our model Hamiltonian is $\hat{H}=\hat{H}_{0}+\hat{H}_{\textrm{imp}}$, 
where $\hat{H}_{0}$ is the Hamiltonian without impurities 
and $\hat{H}_{\textrm{imp}}$ is the impurity Hamiltonian. 
$\hat{H}_{0}$ consists of 
the antiferromagnetic Heisenberg interaction between nearest-neighbor sites
and the magnetic anisotropy, 
\begin{align}
\hat{H}_{0}
=2J\sum\limits_{\langle \boldi,\boldj\rangle}\hat{\boldS}_{\boldi}\cdot \hat{\boldS}_{\boldj}
-D\Bigl[\sum\limits_{\boldi\in A}(\hat{S}^{z}_{\boldi})^{2}
+\sum\limits_{\boldj\in B}(\hat{S}^{z}_{\boldj})^{2}\Bigr],\label{eq:H0}
\end{align}
where 
$\boldi\in A$ and $\boldj\in B$ 
for the $A$ or $B$ sublattice, and 
$\textstyle\sum_{\langle \boldi,\boldj\rangle}=Nz/2$ 
with $N$, as the number of sites, and $z$, as the coordination number; 
the numbers of $A$ and $B$ are equal. 
We assume that $J(>0)$ is much larger than $D(>0)$. 
Then 
we construct $\hat{H}_{\textrm{imp}}$ as follows. 
We first assume that 
one kind of disorder is partial substitution for magnetic ions (see Fig. \ref{fig1}), 
and its main effect is to modify the value of the exchange interaction; 
for simplicity, 
we neglect the disorder effect from the magnetic anisotropy 
because its magnitude will be much smaller. 
Thus 
$\hat{H}_{\textrm{imp}}$ becomes
\begin{align}
\hat{H}_{\textrm{imp}}
=2\sum\limits_{\langle \boldi,\boldj\rangle}\Delta J_{\boldi\boldj}^{(\textrm{imp})}
\hat{\boldS}_{\boldi}\cdot \hat{\boldS}_{\boldj},\label{eq:Himp-0}
\end{align}
with $\Delta J_{\boldi\boldj}^{(\textrm{imp})}=J^{\prime}$ 
for $\boldi\in A_{\textrm{imp}}$, $\boldj\in B_{0}$ 
or for $\boldi\in A_{0}$, $\boldj\in B_{\textrm{imp}}$, 
and $\Delta J_{\boldi\boldj}^{(\textrm{imp})}=J^{\prime\prime}$ 
for $\boldi\in A_{\textrm{imp}}$, $\boldj\in B_{\textrm{imp}}$; 
$A_{0}$ and $B_{0}$ represent $A$ and $B$ sublattice 
for orange circles in Fig. \ref{fig1}(b), 
while $A_{\textrm{imp}}$ and $B_{\textrm{imp}}$ represent 
those for blue ones;  
the numbers of $A_{\textrm{imp}}$ and $B_{\textrm{imp}}$ are equal. 
In a similar way to electron systems~\cite{AGD} 
we suppose that 
impurities are randomly distributed. 
Also, 
we assume that 
$J^{\prime}$ and $J^{\prime\prime}$ are much smaller than $J$. 
Thus 
the main terms of Eq. (\ref{eq:Himp-0}) come from the mean-field-type terms, 
\begin{align}
\hat{H}_{\textrm{imp}}
=-\sum\limits_{\boldi\in A_{\textrm{imp}}}
V_{\textrm{imp}}\hat{S}_{\boldi}^{z}
+\sum\limits_{\boldj\in B_{\textrm{imp}}}
V_{\textrm{imp}}\hat{S}_{\boldj}^{z},\label{eq:Himp}
\end{align}  
where $V_{\textrm{imp}}=
2Sz^{\prime\prime}J^{\prime\prime}$
with $z^{\prime\prime}$, the coordination number for $J+J^{\prime\prime}$. 
Here we have neglected the other mean-field-type terms, 
$-\sum\textstyle_{\boldi\in A}V\hat{S}_{\boldi}^{z}
+\sum\textstyle_{\boldj\in B}V\hat{S}_{\boldj}^{z}$ 
($V=2Sz^{\prime}J^{\prime}$ with $z^{\prime}$, 
the coordination number for $J+J^{\prime}$), 
because those lead to the same effect as the magnetic anisotropy 
in the linear-spin-wave Hamiltonian; 
the effect of the terms in Eq. (\ref{eq:Himp}) 
is different due to the limit of the sum of sites. 

We next express our Hamiltonian in terms of magnon operators. 
For that purpose, 
we use the linear-spin-wave approximation~\cite{Anderson-SW} for a collinear antiferromagnet. 
As a result, 
Eq. (\ref{eq:H0}) becomes 
\begin{align}
\hat{H}_{0}=&
\sum\limits_{\boldq}
\sum\limits_{l,l^{\prime}=A,B}\epsilon_{ll^{\prime}}(\boldq)
\hat{x}^{\dagger}_{\boldq l}\hat{x}_{\boldq l^{\prime}},\label{eq:H0-LSWA}
\end{align}
where $\epsilon_{AA}(\boldq)=\epsilon_{BB}(\boldq)=2S(Jz+D)$
and 
$\epsilon_{AB}(\boldq)=\epsilon_{BA}(\boldq)=2SJ\sum_{j=1}^{z}e^{i\boldq\cdot \boldr_{j}}$, 
and Eq. (\ref{eq:Himp}) becomes
\begin{align}
\hat{H}_{\textrm{imp}}=&
\sum\limits_{\boldq,\boldq^{\prime}}
\sum\limits_{l=A,B}
V^{\textrm{imp}}_{l}(\boldq-\boldq^{\prime})
\hat{x}_{\boldq l}^{\dagger}\hat{x}_{\boldq^{\prime} l},\label{eq:Himp-LSWA}
\end{align}
where $V^{\textrm{imp}}_{l}(\boldQ)
=V_{\textrm{imp}}\frac{2}{N}\sum_{\boldi\in l_{\textrm{imp}}}e^{i\boldQ\cdot \boldi}$. 
Here 
$\sum_{\boldq}$ is the sum of momentum in the first Brillouin zone; 
the magnon operators fulfill $\hat{x}_{\boldq A}=\hat{a}_{\boldq}$ 
and $\hat{x}_{\boldq B}=\hat{b}^{\dagger}_{\boldq}$ 
with 
$\hat{a}_{\boldq}$, the annihilation operator for the $A$ sublattice, 
and $\hat{b}^{\dagger}_{\boldq}$, the creation operator for the $B$ sublattice. 
Then 
we obtain the eigenvalues of Eq. (\ref{eq:H0-LSWA}) 
using the Bogoliubov transformation~\cite{Anderson-SW}: 
$\hat{H}_{0}=\sum_{\boldq}\sum_{\nu=\alpha,\beta}\epsilon_{\boldq}
\hat{x}_{\boldq \nu}^{\dagger}\hat{x}_{\boldq \nu}$, 
where 
$\nu$ is the band index for the $\alpha$ and $\beta$ bands, 
$\epsilon_{\boldq}=\sqrt{\epsilon_{AA}(\boldq)^{2}-\epsilon_{AB}(\boldq)^{2}}$ 
and $\hat{x}_{\boldq l}=\sum_{\nu=\alpha,\beta}U_{l\nu}(\boldq)\hat{x}_{\boldq \nu}$ 
with $U_{A\alpha}(\boldq)=U_{B\beta}(\boldq)=\cosh \theta_{\boldq}$, 
$U_{A\beta}(\boldq)=U_{B\alpha}(\boldq)=-\sinh \theta_{\boldq}$,
and $\tanh 2\theta_{\boldq}=\frac{\epsilon_{AB}(\boldq)}{\epsilon_{AA}(\boldq)}$. 

\textit{Situation.}
As magnon transport in our disordered antiferromagnet, 
we consider $\kappa_{xx}$, 
given by $j_{\textrm{Q}}^{x}=\kappa_{xx}(-\partial_{x} T)$. 
Here $j_{\textrm{Q}}^{x}$ is the thermal current density, 
and $(-\partial_{x} T)$ is the temperature gradient; 
for magnons 
the thermal current is equal to the energy current.  
We focus on the thermal transport rather than the charge transport, 
considered for the localization of electrons~\cite{WeakLoc-review,Nagaoka}, 
because the charge transport is absent in magnets, magnetically ordered insulators.  
Furthermore, 
we consider $\kappa_{xx}$ 
because 
$\kappa_{xx}$ is finite even without external magnetic fields.  
To analyze $\kappa_{xx}$, 
we assume that 
the temperature gradient is so smooth that 
the local equilibrium is reached, that is, the local temperature is definable. 
We also assume that 
the local energy conservation holds. 
Those assumptions are standard ones~\cite{Mahan-text,Luttinger,Oji-Streda,
Tatara}.

\begin{figure}[tb]
\includegraphics[width=84mm]{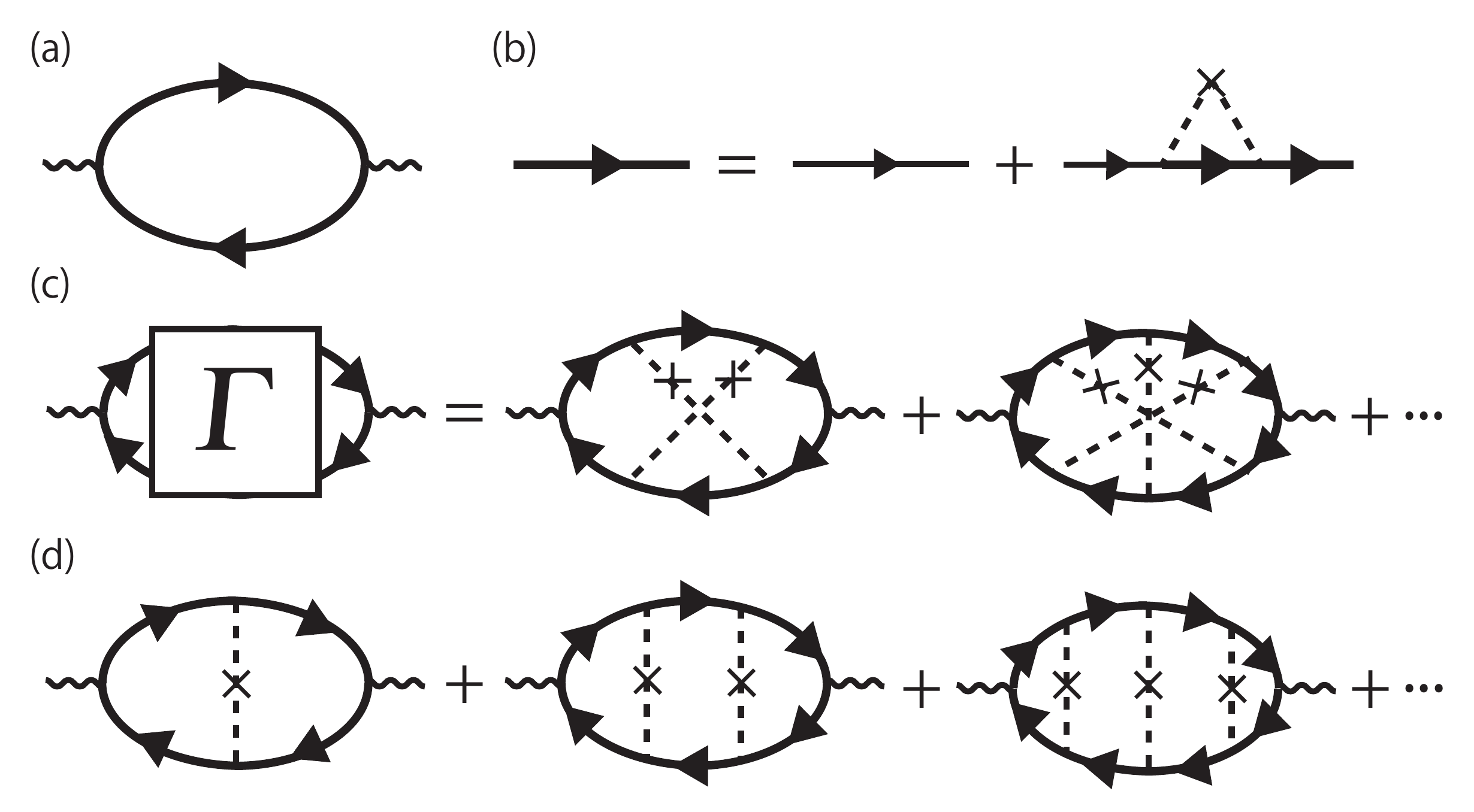}
\caption{
Feynman diagrams of (a) $\kappa_{xx}^{(\textrm{Born})}$, 
(b) the Dyson equation, 
(c) $\Delta\kappa_{xx}$ and 
(d) the contribution from the particle-hole-type vertex corrections. 
Bold arrows and thin arrows denote  
the magnon Green's functions after taking the impurity averaging 
and the magnon Green's functions without impurities; 
a dotted line denotes the impurity scattering. 
}
\label{fig2}
\end{figure}
\textit{Linear-response theory.}
Using the linear-response theory~\cite{Luttinger,Oji-Streda,
Tatara,Eliashberg,NA-Ch,NA-AHESHE}, 
we can express $\kappa_{xx}$ as  
\begin{align}
\kappa_{xx}=\frac{1}{T}
\lim\limits_{\omega\rightarrow 0}
\dfrac{K_{xx}^{(\textrm{R})}(\omega)
-K_{xx}^{(\textrm{R})}(0)}{i\omega},
\end{align}
where $K_{xx}^{(\textrm{R})}(\omega)=K_{xx}(i\Omega_{n}\rightarrow \omega+i0+)$ 
with $\Omega_{n}=2\pi T n$ $(n=0, \pm 1, \pm 2, \cdots)$, 
bosonic Matsubara frequency, 
and 
$K_{xx}(i\Omega_{n})
=\frac{1}{N}
\int^{T^{-1}}_{0}d\tau e^{i\Omega_{n}\tau}
\langle \textrm{T}_{\tau}  
\hat{J}_{\textrm{E}}^{x}(\tau)
\hat{J}_{\textrm{E}}^{x}\rangle$ 
with $T_{\tau}$, a $\tau$-ordering operator~\cite{Mahan-text}.
Since the energy current operator can be derived by 
using the local energy conservation~\cite{Mahan-text}, 
we can derive $\hat{J}_{\textrm{E}}^{x}$ of our model~\cite{Supp}, 
\begin{align}
\hat{J}_{\textrm{E}}^{x}=
\sum\limits_{\boldq}
\sum\limits_{l,l^{\prime}=A,B}
e_{ll^{\prime}}^{x}(\boldq)\hat{x}_{\boldq l}^{\dagger}\hat{x}_{\boldq l^{\prime}},\label{eq:J_E^x}
\end{align}
with $e_{AA}^{x}(\boldq)=-e_{BB}^{x}(\boldq)
=\frac{\partial \epsilon_{AB}(\boldq)}{\partial q_{x}}\epsilon_{AB}(\boldq)$ 
and 
$e_{AB}^{x}(\boldq)=e_{BA}^{x}(\boldq)=\boldzero$. 
Then, 
by using a field theoretical technique~\cite{AGD,Eliashberg,NA-Ch,NA-AHESHE}, 
we obtain
\begin{align}
\kappa_{xx}=&\frac{1}{TN}
\sum\limits_{\boldq,\boldq^{\prime}}
\sum\limits_{\{l_{1}\}}
e_{l_{1}l_{2}}^{x}(\boldq)e_{l_{3}l_{4}}^{x}(\boldq^{\prime})
P\hspace{-4pt}\int^{\infty}_{-\infty}\hspace{-2pt}\frac{d\epsilon}{2\pi}
\Bigl[-\frac{\partial n(\epsilon)}{\partial \epsilon}\Bigr]\notag\\
&\times 
\langle D_{l_{4}l_{1}}^{(\textrm{A})}(\boldq^{\prime},\boldq,\epsilon)
D_{l_{2}l_{3}}^{(\textrm{R})}(\boldq,\boldq^{\prime},\epsilon) \rangle,\label{eq:kappa-start}
\end{align}
where 
$\sum\textstyle_{\{l_{1}\}}\equiv \sum\textstyle_{l_{1},l_{2},l_{3},l_{4}}$, 
$n(\epsilon)$ is the Bose distribution function, 
and $D_{l_{4}l_{1}}^{(\textrm{A})}(\boldq^{\prime},\boldq,\epsilon)$ and 
$D_{l_{2}l_{3}}^{(\textrm{R})}(\boldq,\boldq^{\prime},\epsilon)$,  
the advanced and retarded Green's functions of magnons for $\hat{H}$ 
before taking the impurity averaging. 
(For the derivation, see the Supplemental Material~\cite{Supp}.) 
We have neglected the term including 
$\langle D_{l_{4}l_{1}}^{(\textrm{R})}(\boldq^{\prime},\boldq,\epsilon)
D_{l_{2}l_{3}}^{(\textrm{R})}(\boldq,\boldq^{\prime},\epsilon)\rangle$ 
or $\langle D_{l_{4}l_{1}}^{(\textrm{A})}(\boldq^{\prime},\boldq,\epsilon)
D_{l_{2}l_{3}}^{(\textrm{A})}(\boldq,\boldq^{\prime},\epsilon)\rangle$ 
because the term in Eq. (\ref{eq:kappa-start}) 
is primary in the weak-localization regime~\cite{WeakLoc-review,Nagaoka}.

\textit{Weak-localization theory.}
We 
formulate the weak-localization theory of our disordered antiferromagnet. 
That theory describes 
the disorder effects in the weak-localization regime, 
in which 
the magnitude of $V_{\textrm{imp}}$ is smaller than the magnon energy 
and the impurity concentration, $n_{\textrm{imp}}=\frac{N_{\textrm{imp}}}{N}$, is dilute. 
Since $V_{\textrm{imp}}$ comes from $J^{\prime\prime}$, 
we can apply the perturbation expansion of $\hat{H}_{\textrm{imp}}$ 
to Eq. (\ref{eq:kappa-start}). 
We can employ that expansion 
in a similar way to the longitudinal conductivity of electrons~\cite{WeakLoc-review,Nagaoka} 
and reduce Eq. (\ref{eq:kappa-start}) to 
$\kappa_{xx}=\kappa_{xx}^{(\textrm{Born})}+\Delta\kappa_{xx}$. 
$\kappa_{xx}^{(\textrm{Born})}$ is $\kappa_{xx}$ without vertex corrections 
[Fig. \ref{fig2}(a)], 
\begin{align}
\kappa_{xx}^{(\textrm{Born})}=&\frac{1}{TN}
\sum\limits_{\boldq}
\sum\limits_{\{l_{1}\}}
e_{l_{1}l_{2}}^{x}(\boldq)e_{l_{3}l_{4}}^{x}(\boldq)
P\hspace{-4pt}\int^{\infty}_{-\infty}\hspace{-2pt}\frac{d\epsilon}{2\pi}
\Bigl[-\frac{\partial n(\epsilon)}{\partial \epsilon}\Bigr]\notag\\
&\times 
\bar{D}_{l_{4}l_{1}}^{(\textrm{A})}(\boldq,\epsilon)
\bar{D}_{l_{2}l_{3}}^{(\textrm{R})}(\boldq,\epsilon),\label{eq:kappa^Born}
\end{align} 
and $\Delta\kappa_{xx}$ is 
the contribution from the particle-particle-type vertex corrections [Fig. \ref{fig2}(c)],
\begin{widetext}
\begin{align}
\hspace{-12pt}
\Delta\kappa_{xx}
=&\frac{1}{TN}
\sum\limits_{\boldq,\boldq^{\prime}}
\sum\limits_{\{l_{1}\}}
\sum\limits_{l,l^{\prime}}
e_{l_{1}l_{2}}^{x}(\boldq)e_{l_{3}l_{4}}^{x}(\boldq^{\prime})
P\hspace{-4pt}\int^{\infty}_{-\infty}\hspace{-2pt}\frac{d\epsilon}{2\pi}
\Bigl[-\frac{\partial n(\epsilon)}{\partial \epsilon}\Bigr]
\bar{D}_{l_{4}l^{\prime}}^{(\textrm{A})}(\boldq^{\prime},\epsilon)
\bar{D}_{l_{2}l^{\prime}}^{(\textrm{R})}(\boldq,\epsilon)
\Gamma_{l^{\prime}l}(\boldq+\boldq^{\prime},\epsilon)
\bar{D}_{ll_{1}}^{(\textrm{A})}(\boldq,\epsilon)
\bar{D}_{ll_{3}}^{(\textrm{R})}(\boldq^{\prime},\epsilon).\label{eq:kappa^VC}
\end{align}
\end{widetext}
The contribution from the particle-hole type vertex corrections [Fig. \ref{fig2}(d)] 
will be negligible for our disordered antiferromagnet 
because of the similar argument to electron systems 
with inversion symmetry~\cite{WeakLoc-review,Nagaoka,NA-AHESHE}. 
Then 
the magnon Green's functions 
in Eqs. (\ref{eq:kappa^Born}) and (\ref{eq:kappa^VC}) are determined from 
the Dyson equation [Fig. \ref{fig2}(b)]:
$\bar{D}_{ll^{\prime}}^{(\textrm{R})}(\boldq,\epsilon)
=D_{ll^{\prime}}^{0(\textrm{R})}(\boldq,\epsilon)
+\sum_{l^{\prime\prime}}
D_{ll^{\prime\prime}}^{0(\textrm{R})}(\boldq,\epsilon)
\Sigma_{l^{\prime\prime}}^{(\textrm{R})}(\epsilon)
\bar{D}_{l^{\prime\prime}l^{\prime}}^{(\textrm{R})}(\boldq,\epsilon)$,
where 
$D_{ll^{\prime}}^{0(\textrm{R})}(\boldq,\epsilon)$ is the retarded Green's function without impurities 
and $\Sigma_{l}^{(\textrm{R})}(\epsilon)$ is the retarded self-energy, 
$\Sigma_{l}^{(\textrm{R})}(\epsilon)
=
\gamma_{\textrm{imp}}\sum\textstyle_{\boldq}
\bar{D}_{ll}^{(\textrm{R})}(\boldq,\epsilon)$
with $\gamma_{\textrm{imp}}=\frac{2}{N}n_{\textrm{imp}}V_{\textrm{imp}}^{2}$; 
the advanced quantities are similarly determined.
The vertex function in Eq. (\ref{eq:kappa^VC}) is determined 
from the Bethe-Salpeter equation [Fig. \ref{fig2}(c)]:
$\Gamma_{ll^{\prime}}(\boldQ,\omega)
=\gamma_{\textrm{imp}}^{2}\Pi_{ll^{\prime}}(\boldQ,\omega)
+\sum_{l^{\prime\prime}}
\gamma_{\textrm{imp}}\Pi_{ll^{\prime\prime}}(\boldQ,\omega)
\Gamma_{l^{\prime\prime}l^{\prime}}(\boldQ,\omega)$
with $\Pi_{ll^{\prime}}(\boldQ,\omega)=\sum\textstyle_{\boldq_{1}}
\bar{D}_{ll^{\prime}}^{(\textrm{R})}(\boldq_{1},\omega)
\bar{D}_{ll^{\prime}}^{(\textrm{A})}(\boldQ-\boldq_{1},\omega)$. 

To proceed with the formulation as simple as possible, 
we introduce two simplifications. 
The first one is about the self-energy: 
we consider only the imaginary part. 
This is appropriate because 
its effect is essential 
for the localization~\cite{WeakLoc-review,Nagaoka}.  
The other is about the Green's functions: 
for positive frequencies  
we consider only the positive-pole contribution, 
whereas for negative frequencies 
we consider only the negative-pole contribution. 
For the more precise explanation, 
let us consider $D_{ll^{\prime}}^{0(\textrm{R})}(\boldq,\epsilon)$. 
That for our model is given by
\begin{align}
D_{ll^{\prime}}^{0(\textrm{R})}(\boldq,\epsilon)
=\frac{U_{l\alpha}(\boldq)U_{l^{\prime}\alpha}(\boldq)}{\epsilon-\epsilon_{\boldq}+i\delta}
-\frac{U_{l\beta}(\boldq)U_{l^{\prime}\beta}(\boldq)}{\epsilon+\epsilon_{\boldq}+i\delta},
\label{eq:D^0}
\end{align}
where $\delta\rightarrow 0+$. 
The above first and second terms  provide 
the positive-pole and negative-pole contributions, respectively; 
the first and second terms are dominant for $\epsilon > 0$ and $\epsilon < 0$, 
respectively. 
We thus approximate $D_{ll^{\prime}}^{0(\textrm{R})}(\boldq,\epsilon)$ 
for $\epsilon >0$ by the first term of Eq. (\ref{eq:D^0}), 
and $D_{ll^{\prime}}^{0(\textrm{R})}(\boldq,\epsilon)$ for $\epsilon < 0$ by the second term. 
Combining this and the first simplification with the Dyson equation, 
we obtain
\begin{align}
\bar{D}_{ll^{\prime}}^{(\textrm{R})}(\boldq,\epsilon)
\sim
\begin{cases}
\dfrac{U_{l\alpha}(\boldq)U_{l^{\prime}\alpha}(\boldq)}
{\epsilon-\epsilon_{\boldq}+i\tilde{\gamma}(\epsilon)} 
& (\epsilon > 0),\\[10pt]
-\dfrac{U_{l\beta}(\boldq)U_{l^{\prime}\beta}(\boldq)}
{\epsilon+\epsilon_{\boldq}+i\tilde{\gamma}(-\epsilon)} 
& (\epsilon < 0),
\end{cases}\label{eq:D^R}
\end{align}
where $\tilde{\gamma}(\epsilon)
=(\cosh^{4}\theta_{\boldq}+\sinh^{4}\theta_{\boldq})\gamma(\epsilon)$ 
with $\gamma(\epsilon)=n_{\textrm{imp}}V_{\textrm{imp}}^{2}\pi \rho(\epsilon)$;  
$\rho(\epsilon)$ is the density of states, 
and $\boldq$ of these hypobolic functions are determined by $\epsilon_{\boldq}=|\epsilon|$. 
The advanced quantities are simplified similarly. 

The above simplifications enable us to proceed with the formulation 
in a similar way to the weak localization of electrons~\cite{WeakLoc-review,Nagaoka}. 
First, 
we get a simple expression of $\kappa_{xx}^{(\textrm{Born})}$,
\begin{align}
\kappa_{xx}^{(\textrm{Born})}
\sim 
\frac{1}{TN}
\sum\limits_{\boldq}
\Bigl(
\frac{\partial \epsilon_{\boldq}}{\partial q_{x}}
\epsilon_{\boldq}
\Bigr)^{2}
\Bigl[
-\frac{\partial n(\epsilon_{\boldq})}{\partial \epsilon_{\boldq}}
\Bigr]\tilde{\tau}(\epsilon_{\boldq}),\label{eq:kapp^Born-simpler}
\end{align}
where $\tilde{\tau}(\epsilon_{\boldq})=\tilde{\gamma}(\epsilon_{\boldq})^{-1}$. 
Due to the factor 
$[-\partial n(\epsilon_{\boldq})/\partial \epsilon_{\boldq}]$, 
the contributions for small $q=|\boldq|$ are dominant. 
Then, 
by estimating $\Pi_{ll^{\prime}}(\boldQ,\omega)$ and 
$\Gamma_{ll^{\prime}}(\boldQ,\omega)$ for small $Q=|\boldQ|$, 
we can demonstrate that 
$\Gamma_{ll^{\prime}}(\boldQ,\omega)$ diverges 
in the limit $Q\rightarrow 0$. 
The brief outline of the estimates is as follows (for the details, 
see the Supplemental Material~\cite{Supp}). 
First, 
by using Eq. (\ref{eq:D^R}) and 
performing the momentum sum in $\Pi_{ll^{\prime}}(\boldQ,\omega)$, 
$\Pi_{ll^{\prime}}(\boldQ,\omega)$ for small $Q$ is expressed as
\begin{align}
\hspace{-14pt}
\Pi_{ll^{\prime}}(\boldQ,\omega)
\sim 
\begin{cases}
\frac{u_{l\alpha}^{2}u_{l^{\prime}\alpha}^{2}[1-D_{\textrm{s}}(\omega)Q^{2}\tilde{\tau}(\omega)]}
{\gamma_{\textrm{imp}}(c_{0}^{4}+s_{0}^{4})}
&\hspace{-6pt}(\omega > 0),\\[10pt]
\frac{u_{l\beta}^{2}u_{l^{\prime}\beta}^{2}[1-D_{\textrm{s}}(-\omega)Q^{2}\tilde{\tau}(-\omega)]}
{\gamma_{\textrm{imp}}(c_{0}^{4}+s_{0}^{4})}
&\hspace{-6pt}(\omega < 0),
\end{cases}\label{eq:Pi}
\end{align}
where $u_{l\nu}=U_{l\nu}(\boldq_{0})$,
$c_{0}=\cosh \theta_{\boldq_{0}}$ and $s_{0}=\sinh \theta_{\boldq_{0}}$, 
$\tilde{\tau}(\omega)=\tilde{\gamma}(\omega)^{-1}
=\frac{\tau(\omega)}{(c_{0}^{4}+s_{0}^{4})}$, and 
$D_{\textrm{s}}(\omega)=\frac{1}{4d}
|\frac{\partial \epsilon_{\boldq_{0}}}{\partial \boldq_{0}}|^{2}\tilde{\tau}(\omega)
=\frac{1}{4d}
\boldv_{\boldq_{0}}^{2}\tilde{\tau}(\omega)$, 
the spin-diffusion constant for $d$ dimensions. 
In the above estimate, 
we have approximated the momentum-dependent 
$\cosh^{2}\theta_{\boldq}$ and $\sinh^{2}\theta_{\boldq}$ 
by the typical values, $\cosh^{2}\theta_{\boldq_{0}}$ and $\sinh^{2}\theta_{\boldq_{0}}$; 
$\boldq_{0}$ is a momentum with small magnitude. 
This will be sufficient for a rough estimate 
because 
the dominant contributions come from the terms for small $|\boldq_{1}|$. 
Then, 
combining Eq. (\ref{eq:Pi}) with the Bethe-Salpeter equation, 
we obtain 
\begin{align}
\Gamma_{ll^{\prime}}(\boldQ,\omega)
\sim 
\begin{cases}
u_{l\alpha}^{2}u_{l^{\prime}\alpha}^{2}
\dfrac{\gamma_{\textrm{imp}}}{D_{\textrm{s}}(\omega)Q^{2}\tau(\omega)} 
& (\omega > 0),\\[10pt]
u_{l\beta}^{2}u_{l^{\prime}\beta}^{2}
\dfrac{\gamma_{\textrm{imp}}}{D_{\textrm{s}}(-\omega)Q^{2}\tau(-\omega)} 
& (\omega < 0).
\end{cases}\label{eq:Gamma}
\end{align}
This demonstrates the divergence of $\Gamma_{ll^{\prime}}(\boldQ,\omega)$ 
in the limit $Q\rightarrow 0$. 
This divergence indicates the critical back scattering 
for $\boldq^{\prime}=-\boldq$ in Eq. (\ref{eq:kappa^VC}); 
the other terms about $\boldq^{\prime}$ are nonsingular. 
We thus put $\boldq^{\prime}=-\boldq$ in Eq. (\ref{eq:kappa^VC}) except 
$\Gamma_{l^{\prime}l}(\boldq+\boldq^{\prime},\epsilon)$ 
to estimate the main effect of the critical contribution. 
Under this simplification, 
we can rewrite Eq. (\ref{eq:kappa^VC}) as 
\begin{align}
\Delta \kappa_{xx}
&\sim 
-\frac{1}{TN}
\sum\limits_{\boldq}
\Bigl(
\frac{\partial \epsilon_{\boldq}}{\partial q_{x}}
\epsilon_{\boldq}
\Bigr)^{2}
\Bigl[
-\frac{\partial n(\epsilon_{\boldq})}{\partial \epsilon_{\boldq}}
\Bigr]
\tilde{\tau}(\epsilon_{\boldq})\notag\\
&\times 
\frac{n_{\textrm{imp}}V_{\textrm{imp}}^{2}}{4D_{\textrm{s}}(\epsilon_{\boldq})\gamma(\epsilon_{\boldq})}
\frac{2}{N}\sum\limits_{\boldq^{\prime}}{}^{\prime}
\frac{1}{|\boldq+\boldq^{\prime}|^{2}}.\label{eq:kapp^VC-simpler}
\end{align}
The dominant contributions come from the terms for small $q=|\boldq|$ 
due to the same reason for $\kappa_{xx}^{(\textrm{Born})}$. 
In the sum of $\boldq^{\prime}$ we have replaced 
the lower value of $Q=|\boldq+\boldq^{\prime}|$ by 
a cutoff, $L^{-1}$, which approaches zero in the thermodynamic limit.  
Also, 
we have replaced the upper value of $Q$ by $L_{\textrm{m}}^{-1}$, 
the inverse of the mean-free path.  
(The prime of the sum of $\boldq^{\prime}$ represents those replacements.) 

\textit{Weak localization in a two-dimensional case.}
As a specific example, 
we apply the above theory to 
a two-dimensional case on the square lattice for arbitrary $S$. 
In this case, 
$\epsilon_{ll^{\prime}}(\boldq)$ are 
$\epsilon_{AA}(\boldq)=\epsilon_{BB}(\boldq)=2S(4J+D)$ 
and $\epsilon_{AB}(\boldq)=\epsilon_{BA}(\boldq)=4SJ(\cos q_{x}+\cos q_{y})$. 
Since we have 
$\frac{2}{N}\sum\textstyle_{\boldq^{\prime}}^{\prime}|\boldq+\boldq^{\prime}|^{-2}
=\int_{L^{-1}}^{L_{\textrm{m}}^{-1}}\frac{d Q}{2\pi}Q Q^{-2}
=\frac{1}{2\pi}\ln(\frac{L}{L_{\textrm{m}}})$ 
and we can approximate $\gamma(\epsilon_{\boldq})$ and 
$D_{\textrm{s}}(\epsilon_{\boldq})$ in Eq. (\ref{eq:kapp^VC-simpler}) 
by $\gamma_{0}=\gamma(\epsilon_{\boldq_{0}})$ and 
$D_{\textrm{s}0}=D_{\textrm{s}}(\epsilon_{\boldq_{0}})$, respectively,  
$\kappa_{xx}=\kappa_{xx}^{(\textrm{Born})}+\Delta\kappa_{xx}$ is reduced to 
\begin{align}
\kappa_{xx}
&=\kappa_{xx}^{(\textrm{Born})}
\Bigl[
1-\frac{n_{\textrm{imp}}V_{\textrm{imp}}^{2}}{[\pi\boldv_{\boldq_{0}}^{2}/(c_{0}^{4}+s_{0}^{4})]}
\ln \Bigl(\frac{L}{L_{\textrm{m}}}\Bigr)
\Bigr].\label{eq:kappa_xx-last}
\end{align}
This shows that 
the critical back scattering causes the logarithmic suppression, 
which diverges in the thermodynamic limit. 
Thus 
magnons are localized at low temperatures 
in the two-dimensional disordered antiferromagnet. 

The above $\ln L$ dependence may indicate that 
the $\ln T$ dependence emerges in the presence of inelastic scattering 
because of a similar argument to the electron system~\cite{Thouless-Inela,Anderson-Inela}. 
We have considered only the elastic scattering of $\hat{H}_{\textrm{imp}}$. 
However, 
if we consider the interaction between magnons, 
it causes the inelastic scattering, resulting in 
a temperature-dependent mean-free path. 
Since that is expressed as a power function of $T$, 
the $\ln L$ dependence of $\kappa_{xx}$ may result in 
the $\ln T$ dependence in the presence of the inelastic scattering.

\textit{Discussion.}
We first discuss the validity of our theory. 
It treats partial substitution for magnetic ions as impurities, 
and analyzes the effect on $\kappa_{xx}$ in the weak-localization regime. 
Such a situation may be realized 
by substituting some of the magnetic ions with different ones, 
which belong to the same family of the periodic table; 
an example is the substitution of Ag ions for Cu ions. 
We have considered such partial substitution because 
magnetic ions in the same family have the same $S$ 
due to the same number of electrons in the open shell 
[e.g., in La$_{2}$Cu$_{1-x}$Ag$_{x}$O$_{4}$, $(3d)^{9}$ for Cu ions 
and $(4d)^{9}$ for Ag ions], 
and because its main effect is to change the exchange interaction. 
Then, 
our theory is applicable to disordered Heisenberg antiferromagnets 
for any $S$ and any dimension, 
whereas the specific example considered here is the two-dimensional case. 
Since our theory uses the linear-spin-wave approximation, 
which can be appropriate at low temperatures, 
our theory generally can describe the weak localization of magnons 
of any disordered Heisenberg antiferromagnets at low temperatures. 
In our theory 
the temperature effect comes from the Bose distribution function. 

We now turn to experimental implications. 
Our main result shows that 
the magnon energy current parallel to the temperature gradient 
is suppressed drastically in the disordered two-dimensional antiferromagnet. 
This property is experimentally testable 
by measuring and comparing $\kappa_{xx}$ 
in cases without and with partial substitution of magnetic ions; 
for example, 
this can be performed in a quasi-two-dimensional antiferromagnet, 
such as La$_{2}$Cu$_{1-x}$Ag$_{x}$O$_{4}$. 
In addition, 
this property will be useful for a thermal switch as a spintronics device 
because turning the magnon thermal current 
on and off is controllable by partial substitution for the magnetic ions. 

Our theory also has several theoretical implications. 
Our theory may provide a starting point 
for further studies of magnon localization 
because the weak-localization theory~\cite{RG-4persons,Gorkov} for electrons 
under time-reversal symmetry 
opened up further research in various situations~\cite{WeakLoc-review,Nagaoka}. 
In particular, 
by using or extending our theory, 
it is possible to understand 
how the dimension of the system and the symmetry of the Hamiltonians 
affect the weak localization of magnons in disordered antiferromagnets. 
Furthermore, 
in a similar way to our theory, 
we can construct the weak-localization theory of magnons 
for another magnet 
even if its Hamiltonian includes more complex terms. 
That study may help understand 
the difference due to the magnetic structure 
and exchange interactions. 

\textit{Summary.}
We have formulated $\kappa_{xx}$ 
of the disordered Heisenberg antiferromagnet in the weak-localization regime, 
and showed the weak localization of magnons in two dimensions. 
This theory is valid at low temperatures for any $S$ and any dimension. 
We have shown that 
the multiple impurity scattering critically enhances the back scattering of magnons, 
resulting in the logarithmic suppression of $\kappa_{xx}$ with $L$ in two dimensions. 
Also, 
we have argued that 
this logarithmic suppression may result in the logarithmic temperature dependence of $\kappa_{xx}$ 
due to the inelastic scattering. 
Our weak localization can be observed experimentally by 
measuring $\kappa_{xx}$ in a quasi-two-dimensional antiferromagnet, 
such as La$_{2}$Cu$_{1-x}$Ag$_{x}$O$_{4}$. 
Furthermore, 
our weak localization may be utilized as a thermal switch. 
This work provides a starting point for further research of the weak localization of magnons. 

\begin{acknowledgments}
This work was supported by CREST, JST and Grant-in-Aid 
for Scientific Research (A) (17H01052) from MEXT, Japan.
\end{acknowledgments}


\end{document}